# Novel electronic and magnetic properties of BN sheet decorated with hydrogen and fluorine

Jian Zhou [1], Qian Wang [2], Qiang Sun [1], and Puru Jena [2]

[1]Department of Advanced Materials and Nanotechnology, and Center for Applied Physics and Technology, Peking University, Beijing 100871, China

[2]Department of Physics, Virginia Commonwealth University, Richmond, VA 23284

(Oct. 2, 2009)

## Abstract

First principles calculations based on density functional theory reveal some unusual properties of BN sheet functionalized with hydrogen and fluorine. These properties differ from those of similarly functionalized graphene even though both share the same honeycomb structure. (1) Unlike graphene which undergoes a metal to insulator transition when fully hydrogenated, the band gap of the BN sheet significantly narrows when fully saturated with hydrogen. Furthermore, the band gap of the BN sheet can be tuned from 4.7 eV to 0.6 eV and the system can be a direct or an indirect semiconductor or even a half-metal depending upon surface coverage. (2) Unlike graphene, BN sheet has hetero-atomic composition, when co-decorated with H and F, it can lead to anisotropic structures with rich electronic and magnetic properties. (3) Unlike graphene, BN sheets can be made ferromagnetic, antiferromagnetic, or magnetically degenerate depending upon how the surface is functionalized. (4) The stability of magnetic coupling of functionalized BN sheet can be further modulated by applying external strain. Our study highlights the potential of functionalized BN sheets for novel applications.

## Introduction

Dimensionality has become an important parameter to tune the properties of materials. Currently tremendous effort has been devoted to two dimensional (2D) atomic layer based materials. One of such example is graphene[1] which has many advantages compared to carbon nanotubes (CNTs) when used in polymeric composites. The covalently bonded honeycomb lattice of the graphene sheet shows high stability and exceptional properties[2] such as high electron mobility even at room temperature, ambipolar effect, Klein tunneling, and anomalous quantum hall effect. Especially, electrons in graphene sheet behave like mass-less Dirac fermions making the observation of several relativistic effects possible. The impressive progress in graphene research has motivated scientists to explore other 2D atomic based materials. Among them BN sheet has become a hotly pursued system as it shares the same honeycomb lattice structure as graphene. Using a chemical-solution-derived method starting from single-crystalline hexagonal boron nitride, Han and co-workers have successfully synthesized BN mono-atomic layer.[3] Very recently, an efficient method to fabricate high-yield 2D BN sheets has been developed by using a sonication–centrifugation technique[4] where milligram quantity of BN sheets are achievable, and ultimately pure BN sheets can be obtained based on a highly pure precursor. The sheet thickness can be adjusted by the centrifugation speed.

The advances in experimental synthesis of BN sheet have led us to explore the properties of BN sheet by decorating its surface. Although BN sheet has similar geometry as graphene sheet, it can display different properties. For example, unlike graphite BN layers are stable under high temperature up to 1000 K. BN sheet is semiconducting while graphene sheet is metallic. In order to open a band gap in graphene, surface modification such as hydrogenation is needed. The resulting graphane sheet[5,6] has a band gap of 4.5 eV which is tunable by modifying its surface with atoms such as F.[7] Similar decoration of BN sheet raises some interesting questions: (1) since the BN sheet is already semiconducting with a band gap of 4.7 eV, how would its band gap change when hydrogenated or fluorinated? (2) When half of the H atoms are removed from a fully hydrogenated graphene sheet (graphane), the resulting semi-hydrogenated graphene sheet (graphone) becomes ferromagnetic,[8] but it is still nonmagnetic when semi-fluorinated. What would be the magnetic properties of semi-hydrogenated or semi-fluorinated BN sheet? Is it ferromagnetic, antiferromagnetic, or nonmagnetic? (3) All C atoms in graphene sheet are equivalent with covalent bonding between them. In contrast B and N sites in BN sheet are not equivalent. The charge transfer from B to N permit the bonding between them to be more ionic, thus the properties of semi decorated BN sheet will depend upon which sites are decorated with H and/or F. How would co-decoration of BN sheet with H on one side and F on the other side affect its electronic and magnetic properties?

In this paper, we have made extensive studies of the electronic structures and magnetic properties of fully- and semi-decorated BN sheet using H and/or F atoms. We show that for a fully hydrogenated BN sheet, the energy band gap is reduced from that in pristine BN sheet. With semi-decoration on different sites, BN sheet can display ferromagnetic (FM) or antiferromagnetic (AF) properties. And for semi-fluorinated BN sheet, contrary to what has been found in the case of BN nanotube, FM and AF states are nearly degenerated. In addition, the relative stability of FM and AF states can be further modulated by applying external strain.

## METHODS

Our calculations are based on spin-polarized density functional theory (DFT) using generalized gradient approximation (GGA)[9] for exchange-correlation potential. We have used Perdew-Burke-Ernzerhof (PBE) functional for GGA as implemented in the Vienna *ab initio* Simulation Package (VASP).[10,11] For the geometric and electronic structural calculations, a supercell consisting of four-fold unit cells of *h*-BN sheet is used with a vacuum space of 15 Å between two layers to avoid interactions between them. Pseudopotentials with $2s^2 2p^1$, $2s^2 2p^3$, $1s^1$, and $2s^2 2p^5$ valence electron configurations respectively for B, N, H, and F atoms are used. The Brillouin zone is represented by Monkhorst-Pack special k-point mesh[12] of 7x7x1. The energy cutoffs, convergence in energy and force are set to 400 eV, $1\times10^{-4}$ eV, and 0.01 eV/Å, respectively. Optimizations are performed using conjugated gradient method and without any symmetric constraints. The accuracy of our calculation procedure is tested using pristine *h*-BN sheet. The optimized bond length of B-N of 1.446 Å is in good agreement with experimental value of 1.45 Å. The calculated band gap of BN sheet of 4.71 eV (Fig. 1) is also in good accordance with previous theoretical result.[13]

## Results and Discussion

In the following we discuss the electronic structure, and properties of BN sheet when (1) fully hydrogenated, (2) semi-hydrogenated, (3) semi-fluorinated, (4) external strain applied, and (5) semi-hydrogenated and semi- fluorinated.

### *1. Fully hydrogenated BN sheet*

We first discuss the results for a fully hydrogenated *h*-BN sheet (labeled as H-BN-H) which is very similar in geometry to its carbon counterpart, graphane, as shown in Fig. 1a. After hydrogenation, B and N atoms become $sp^3$ hybridized which distorts the planar geometry forming a zigzag configuration as C atoms do in graphane[6]. The distance between B and N planes is found to be 0.518 Å. The hydrogen atoms are adsorbed on the top site of B and N atoms with H-B and H-N bond lengths of 1.201 Å and 1.036 Å, respectively. The bond length of H-B is a little larger than that of H-N bond length because of the difference in bonding. Mulliken analysis suggests that H atoms adsorbed on B and N are both positively charged carrying charges of 0.012 and 0.129 electrons, respectively. The energy band structure and partial density of state (PDOS) of H-BN-H are plotted in Fig. 1b. It is found to be non-magnetic with a direct band gap of 3.33 eV, which is lower than that in the pristine BN sheet, namely, 4.71 eV. The valence band maximum (VBM) and conduction band minimum (CBM) are both located at the Γ point in the reciprocal space. Here we see some differences between fully hydrogenated graphene and BN sheet. Hydrogenation opens a band gap in the former while it reduces the band gap in the latter. In fact, introduction of two H atoms introduces two occupied energy bands (one in majority and one in minority band sharing the same energy). These are located just above the valence band of pristine BN sheet. Accordingly, the band gap of the system is reduced.

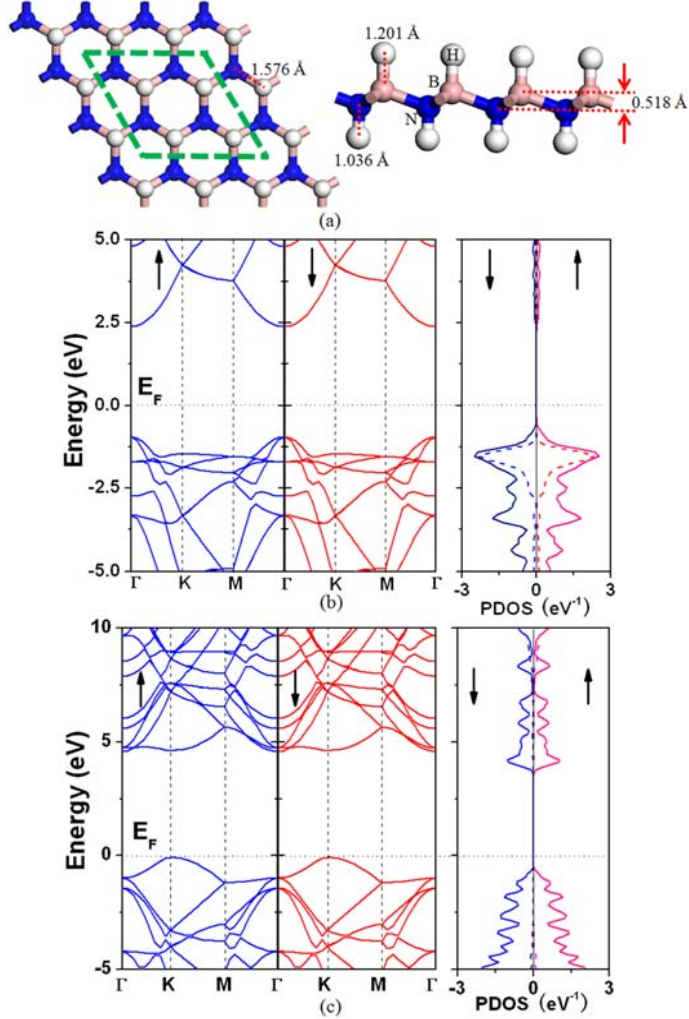

*Figure 1* (a) Optimized structure of hydrogenated BN nanosheet, H-BN-H. The dashed rhombus outlines the supercell used in our calculation. (b) and (c) are calculated band structure and PDOS of H-BN-H and pristine BN sheet, respectively. In PDOS, dashed and solid curves correspond to *s* and *p* orbitals, respectively.

## 2. *Semi-hydrogenated BN sheet*

In graphene sheet all the C sites are equivalent. Thus, when removing half of H atoms from a fully hydrogenated graphene sheet (graphane), we have only one option for the semihydrogenated sheet[8]. However, in BN sheet, B and N sites are not equivalent and semi-hydrogenation can be accomplished by removing H from either B sites or N sites. Which is the preferred configuration? To find this we first considered semi-hydrogenation by placing H on B sites, as shown in Fig. 2. We label this case as H-BN.

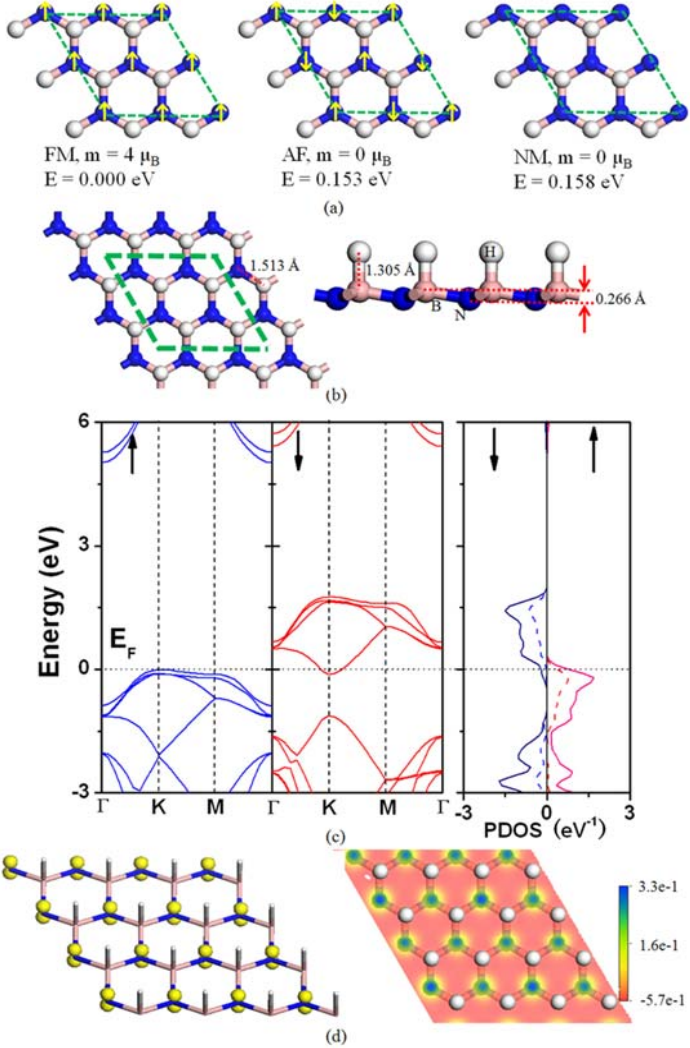

*Figure 2* Calculated results of H-BN. (a) three magnetic coupling states, (b) optimized structure of the FM state, (c) calculated band structure and PDOS of H-BN in FM state, (d) iso-surface (0.5 electron/Å$^3$) and 2D slice of spin density ($\rho_\alpha$-$\rho_\beta$) in FM state.

Here the N atoms remain exposed. The geometric optimization showed that the BN surface became more planar than that of H-BN-H, and the distance between B plane and N plane is reduced to 0.266 Å. The H-B bond length is elongated to 1.305 Å. These results are similar to those found in graphane and graphone[8]. In H-BN configuration, we found that the N atoms possess about 1 $\mu_B$ magnetic moment, while B and H atoms carry very small magnetic moments. To verify that the system is indeed magnetic we calculated the total energies for three magnetic coupling configurations between N atoms (Fig. 2a): (1) FM coupling, (2) AF coupling and (3) non-magnetic (NM) coupling. The results showed that the FM coupling between N atoms is energetically lower than that in AF and NM configurations by 0.153 eV and 0.158 eV, respectively. Since our calculated supercell is based on four unit cells of BN sheet, the energy differences are thus 38 meV and 40 meV per unit cell, respectively. This indicates that the semi-hydrogenated H-BN nanostructure exhibits FM coupling. The physics involved can be described as the following: In pristine BN sheet, the charge transfer from B to N and the orbital hybridization make electrons paired and the system is nonmagnetic. The $2p_z$ electrons of N atoms contribute to the highest occupied VBM while $2p_z$ electrons of B atoms contribute to lowest unoccupied CBM. When semi-hydrogenated on B sites, B atoms are covalently bonded with H atoms forming sp$^3$ hybridization and almost no charge transfer occurs from B to N. Thus $2p_z$ electrons on N atoms remain unpaired. The extended p-p interaction results in a long-range magnetic coupling between $2p$ moments as found in graphone sheet[8]. Conventionally, magnetic moment comes from unfilled *d* or *f* orbitals of metallic atoms. Using mean field theory and the energy difference between FM and AF state, we can estimate the Curie temperature of H-BN to be 293 K and 440 K, depending upon whether the system is treated as 3D and 2D, respectively.

Band structure and PDOS are plotted in Fig. 2c. We find that the system is half-metallic where majority spin state remains as an indirect band gap insulator with band gap of 5.05 eV, while minority spin state is metallic. To visualize the magnetic property clearly, we plotted the iso-surface and a two-dimensional slice of spin density (Fig. 2d). It can be seen that the magnetism on N atoms are mainly contributed by unsaturated $2p_z$ orbital, which is consistent with the results obtained from PDOS. We find that one H atom adsorbed on B site introduces one occupied energy band in majority channel as well as one unoccupied energy band with slightly higher energy in the minority channel. The two energy bands are located slightly above the highest occupied valence band of pristine BN sheet. If the concentration of H atoms is increased to the level of semi-hydrogenation, the new induced occupied and unoccupied energy bands will be broadened due to hybridization effect. They will collapse with each other, and give rise to half-metallic band structure.

Next we consider semi-hydrogenation on N sites (denoted as H-NB, see Fig. 3). The optimized H-NB sheet becomes less planar than H-BN sheet discussed above. The distance between B and N planes is calculated to be 0.450 Å. The H-N bond length of 1.083 Å is again a little longer than that in H-BN-H, namely, 1.036 Å. Since B atom has a valence configuration of $2s^2 2p^1$, once N sites are terminated with H, the charge transfer from B to N is diminished, leaving 2p electron spin in B site unpaired. In fact, we found that the un-hydrogenated B atoms carried a magnetic moment of about $0.75\mu_B$, while no net magnetic moment was found on H or N atoms. We computed total energies for three magnetic configurations (FM, AF, and NM). The AF state is found to be energetically most stable. The FM and NM states are higher in energy than that of AF state by 0.476 eV and 0.500 eV, respectively, corresponding to 119 meV and 125 meV per unit cell. These energy differences include magnetic contribution from both exchange interaction as well as from changes in geometry. We found that in FM state the H-N bond length is 1.071 Å, which is shorter than that in the AF state. The energy of the H-NB in AF state is higher than its isomer H-BN in its FM state by 1.36 eV, suggesting that H-NB configuration is energetically less stable than H-BN configuration.

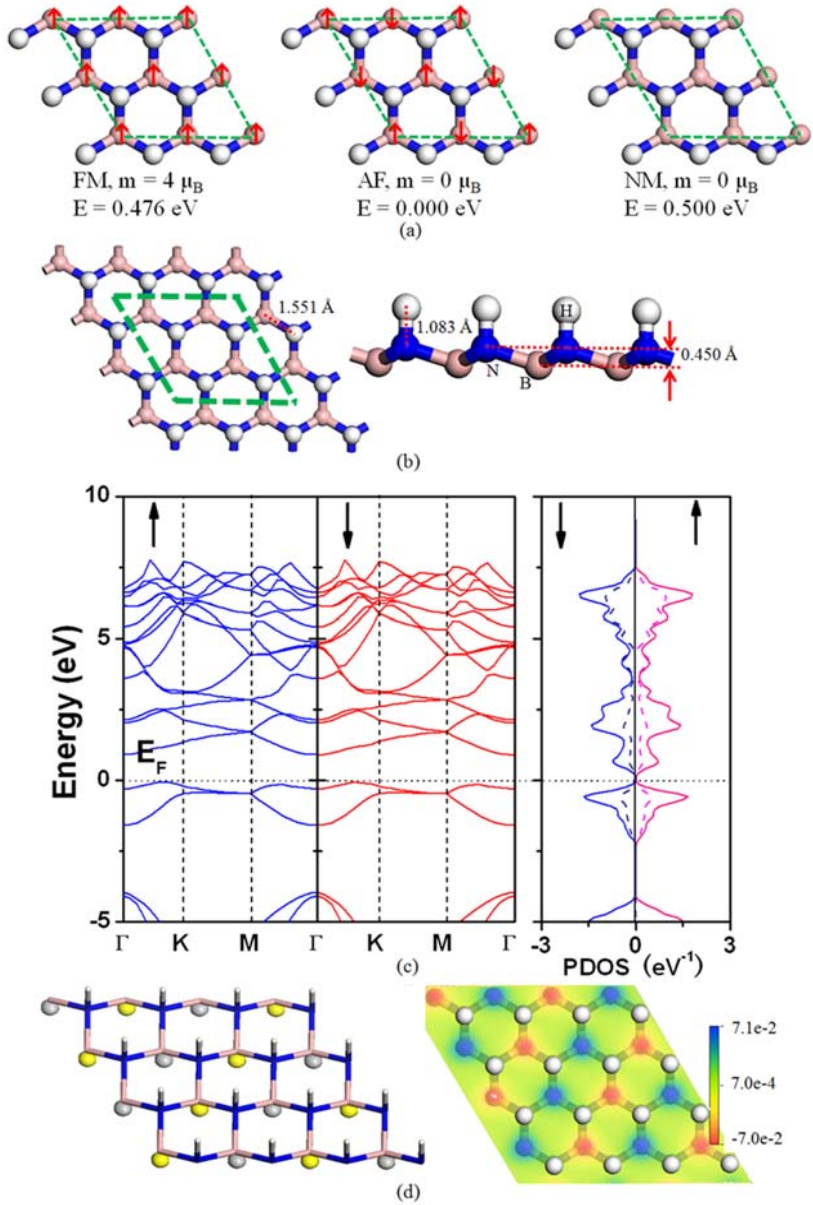

*Figure 3* The legend is similar to that in Fig. 2 while results in (b), (c) and (d) are for the AF state. (d) The yellow and gray indicate the positive and negative values, respectively in the iso-surface (0.38 electron/Å$^3$).

The band structure and PDOS for the AF state of H-NB are plotted in Fig. 3c, showing that the system is an indirect band gap semiconductor where band gap is reduced to 0.97 eV. Detailed analysis of PDOS reveals again that the magnetic moment on B atoms is due to the unsaturated 2$p$ orbital, which is clearly shown in Fig. 3d. The system with H atoms adsorbed on N atoms of BN sheet in AF state also shows that some occupied (majority) and unoccupied (minority) energy bands are newly formed just under the conduction band (related to the electron occupation in the lowest unoccupied conduction band arising from B $2p$ orbital of pristine BN sheet), which makes the system *n*-type semiconductor.

### *3. Semi- fluorinated BN sheet*

From above we have seen that the electronic and magnetic properties of BN sheet can be tuned by decorating B or N sites with H atoms. On the other hand, fluorine atoms decorated on nanostructures such as graphene,[7] carbon nanotube (CNT),[14-16] fullerene ($C_{60}F_{18}$,[17,18] $C_{60}F_{36}$,[19,20] $C_{60}F_{48}$,[20-22] $C_{60}F_{60}$,[23] $C_{58}F_{18}$[24]) have also been studied experimentally and theoretically. Kudin *et al.*[10] showed that fluorine adsorptions on different sites can change CNT from metallic to semiconducting. In this section, we discuss the effect of F atoms decorated on the BN sheet and its corresponding electronic and magnetic properties.

Due to weak binding between F and N atoms, we only consider semi-fluorination of BN sheet in which F atoms attached to B atoms in BN (termed as F-BN) (Fig. 4). In the optimized geometric structures, B and N atoms form planar configurations. The distance between B and N planes is found to be 0.375 Å, and F-B bond length is 1.415 Å. Again, we found that the N atoms carry a magnetic moment of about $1\mu_B$. Three magnetic states with their corresponding energies are given in Fig. 4a, showing that the AF state is the most stable one and lies 0.027 eV and 1.114 eV lower than FM and NM states, respectively. Per unit cell, the energy difference between AF and FM state is thus 7 meV. This energy is small and comparable to that due to thermal fluctuation. Hence, we conclude that the FM and AF states are nearly degenerated. Their corresponding band structure and PDOS are plotted in Fig. 4. We can see that that FM state is half-metallic, where the majority channel remains insulating with a band gap of 6.57 eV, and the minority channel is conducting. Analysis of PDOS shows that the magnetism on N atoms is mainly contributed from the unpaired $2p_z$ orbital, which can be visualized from the iso-surface of spin density (Fig. 4e). The relevant mechanism is similar to what we discussed in H-BN sheet. The binding of F atoms on B sites make the 2p electrons on N sites unpaired. We also found that when two F atoms are introduced, two unoccupied energy bands are induced in the minority spin channel just slightly above the valence band. When more F atoms are adsorbed, the system finally becomes half-metallic.

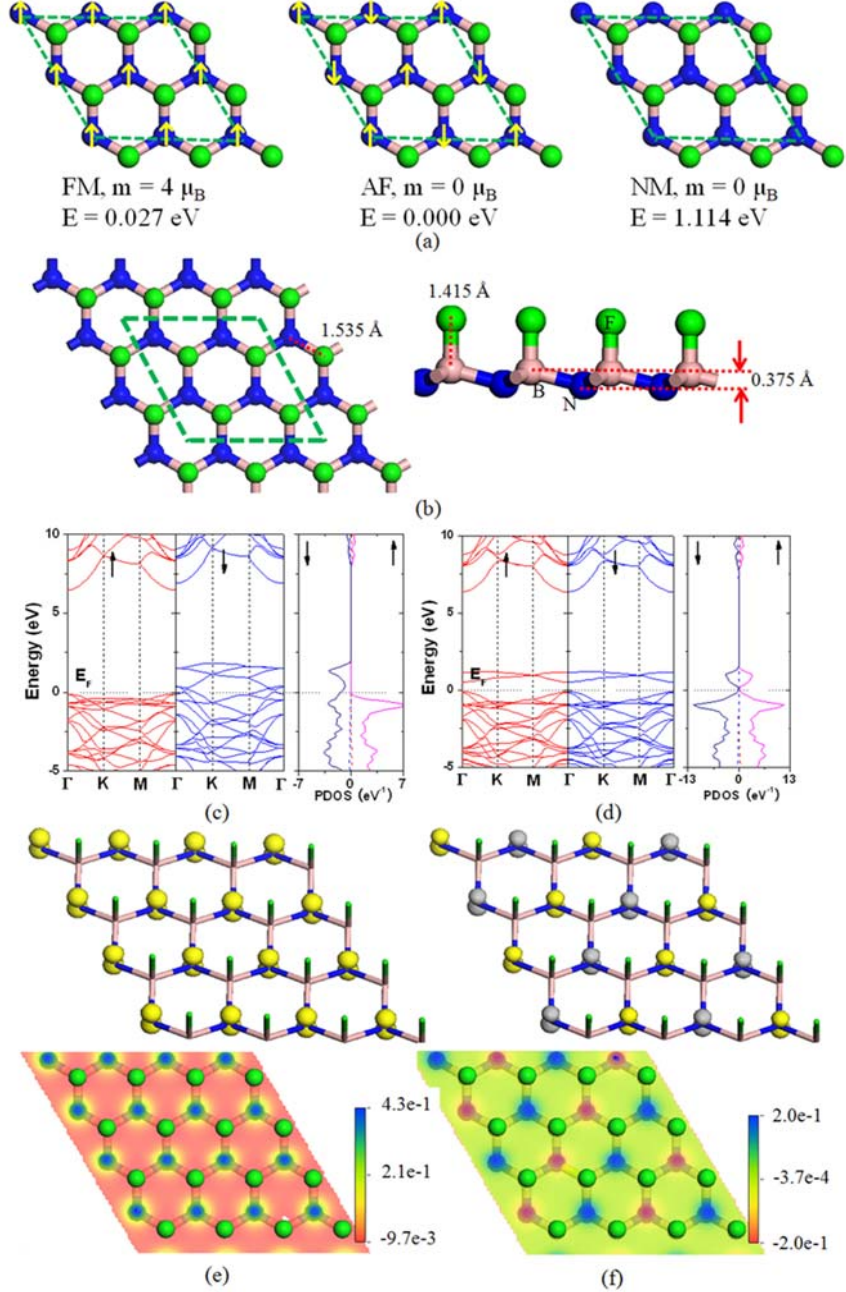

*Figure 4* Calculated results of F-BN. (a) relative energies for three magnetic coupling. (b) Structure of F-BN in AF state. (c) and (e) calculated band structure, PDOS, as well as iso-surface (0.5 electron/Å$^3$) and 2D slice of spin density of F-BN in FM state. (d) and (f) are same as (c) and (e) but for AF state.

For the AF state, the system becomes a direct band gap semiconductor with a small band gap of 0.63 eV. Just above the valence band of pristine BN sheet, we note that there are four induced energy bands in the conduction manifold (two in majority and two in minority spin channel). A large band gap exists above these two bands. The iso-surface and two-dimensional slice of spin density are plotted in Fig. 4f.

### *4. Applied stress on Semi- fluorinated BN sheet*

We found that magnetic properties of fluorinated BN sheet are different from that in the fluorinated BN nanotube.[25] The latter has been found to be FM. To see if this difference is due to strain caused by curvature effects on the BN nanotube, we applied in-plane tensile and compressive stress in the fluorinated BN sheet. For the F-BN system under tensile stress, the energy difference between AF and FM state increases with higher tension, and AF is always more stable than FM state. But for the compressive stress on F-BN, we found that the energy of FM state increases slower than AF state, and at 5% in-plane compression, FM state is energetically more stable than AF state by 32 meV per unit cell.

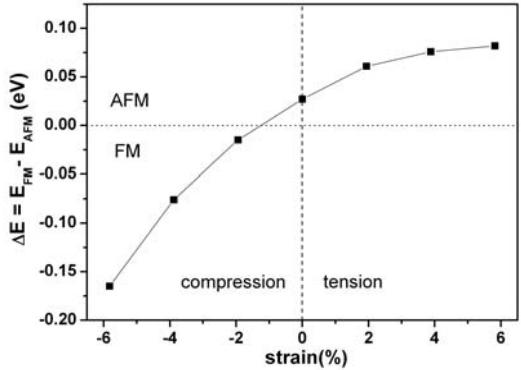

*Figure 5*.  Stability of magnetic state as a function of strain.

### *5. BN sheet co-decorated with H and F*

In this section, we study the effect of co-decoration of the BN sheet with both H and F and its effect on the electronic and magnetic properties. This is achieved by decorating one side of the BN sheet with F and the other side with H. The F atoms are attached to the B site while H atoms are attached to the N sites. We label this system as F-BN-H (Fig. 6). The optimized F-B and H-N distances are shorter than those in F-BN and H-NB systems indicating that the bonding is stronger than that in the systems discussed above. Consequently, the distortion of the planar BN sheet is larger than that in the F-BN and H-NB systems as can be seen from the distance between B and N layers, which is 0.548 Å. The co-decorated F-BN-H system is a direct band gap semiconductor with a band gap of 2.55 eV which is smaller than those of pristine BN, or H-BN-H, but much larger than those in H-NB (AF) and F-BN (AF) sheets.

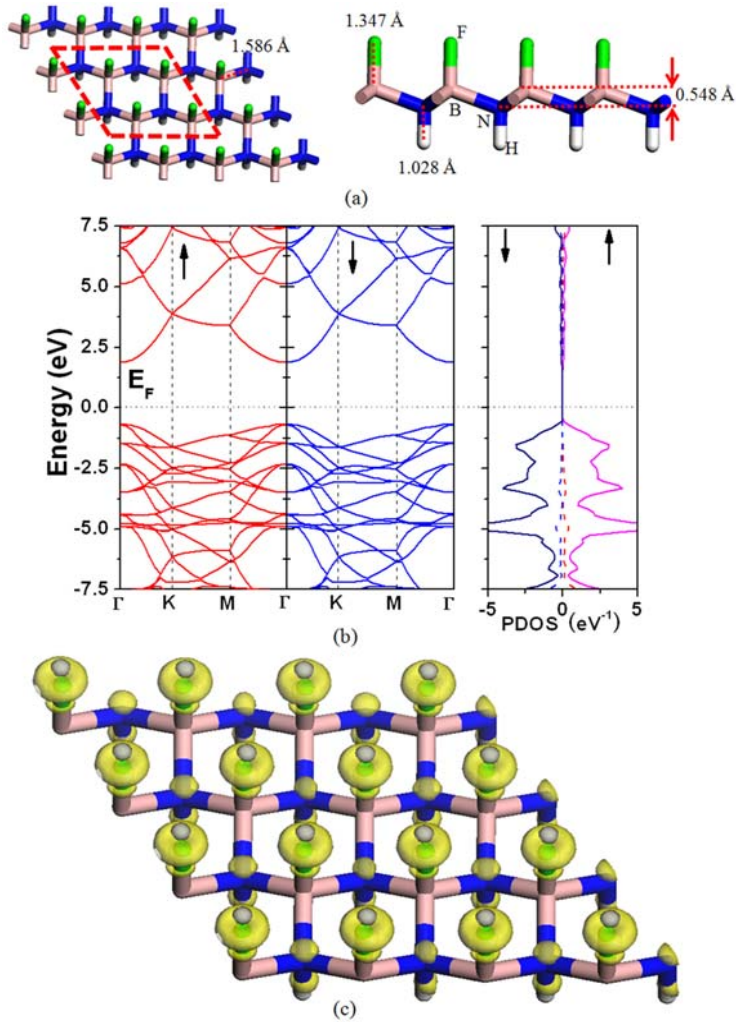

*Figure 6* (a) Calculated geometric structure, (b) band structure and PDOS of F-BN-H, and (c) the charge density difference ($\rho_{F-BN-H}-\rho_F-\rho_{BN}-\rho_H$) of F-BN-H.

What is more interesting is that co-decoration using H and F atoms induces anisotropy in the system. The pristine BN sheet is a two dimensional isotropic system. When decorated with F atom on B sites and H atom on N sites, the large difference in the electron affinity of H and F makes the charge distribution asymmetric; F atoms receive electrons and become negatively charged on one side while H atoms donate electrons and become positively charged on the other side. This asymmetric charge distribution induced by the co-decoration in BN sheet can be visualized from the electron density difference ($\rho_{F-BN-H}-\rho_F-\rho_{BN}-\rho_H$) shown in Fig. 6(c). One can take advantage of this asymmetric charge distribution to promote self-assembly under an applied electric field. Thus, the weakness of van der Waals interactions between the pristine BN sheets can be overcome and synthesis of more stable BN sheets by using linkers may be possible. There is a growing interest in introducing anisotropy[26,27] to nanostructures for novel applications in molecular recognition, self-assembly, photonic crystals, sensors, drug delivery, surfactants, solar cells, and display materials. We hope that the co-decorated anisotropic F-BN-H sheet may have potential applications.

Table I. Summary of electronic and magnetic properties of the systems studied. Symbol *s/h* refers to semiconductor or half-metal, and *d/i* refers to direct or indirect band gap of semiconductor. $E_g$ is the energy band gap.

|  | BN | H-BN-H | H-BN (FM) | H-NB (AF) | F-BN (FM) | F-BN (AF) | F-BN-H |
|---|---|---|---|---|---|---|---|
| s/h | s | s | h | s | h | s | s |
| d/i | d | d | - | i | - | d | d |
| $E_g$/eV | 4.71 | 3.33 | - | 0.97 | - | 0.63 | 2.55 |

**Conclusions**

We have systematically studied the modulation of electronic and magnetic properties through hydrogenation and/or fluorination of BN sheet. In Table 1 we summarized the results. The main conclusions are as follows: (1) For a fully decorated system, our calculations show that they are all direct band gap semiconductors. (2) In semi-decorated systems, H-BN, F-BN and H-NB, the unsaturated N or B atoms carry magnetic moments which are contributed mainly by $2p_z$ orbital. H-BN and H-NB systems couple respectively ferromagnetically and antiferromagnetically. However, in F-BN system the ferromagnetic and antiferromagnetic states are energetically nearly degenerate. (3) The induced energy band is located above the valence band or below the conduction band and reflects how electrons are transferred. For the H-BN and F-BN in FM state, both the systems possess half-metallic property with majority spin channel being semiconducting while minority spin channel being metallic. These are related to the extension of occupied and unoccupied energy bands above the valence band of pristine BN sheet. The H-NB and F-BN systems in AF state, behave as semiconductors with small energy gaps. (4) The in-plane strain can be used to tune the relative stability of FM and AF configurations of F-BN system. (5) Co-decoration with H and F atoms can introduce anisotropy in charge distribution, which can be used to facilitate self-assembly under an applied electric field making it possible to overcome the weakness of van der Waals interactions between the pristine BN sheets. (6) All surface decorations of the BN sheet reduce the energy band gap displaying different behaviors from that in the graphene sheet. The diverse properties of decorated BN sheets have the potential for wider applications of 2D-based materials and devices.

**Acknowledgment**

This work is partially supported by grants from the National Natural Science Foundation of China (NSFC-10744006, NSFC-10874007) and from the US Department of Energy.

**REFERNECES**

1. K. S. Novoselov, A. K. Geim, S. V. Morozov, D. Jiang, Y. Zhang, S. V. Dubonos, I. V. Grigorieva, A. A. Firsov, Science **306**, 666 (2004).

2. A. K. Geim, Science **324**, 1530 (2009).

3. W. Q. Han, L. Wu, Y. Zhu, K. Watanabe, T. Taniguchi, Appl. Phys. Lett. **93**, 223103 (2008).

4. C. Zhi, Y. Bando, C. Tang, H. Kuwahara, D. Golberg, Adv. Mater. **21**, 2889 (2009).

5. D. C. Elias, R. R. Nair, T. M. G. Mohiuddin, S. V. Morozov, P. Blake, M. P. Halsall, A. C. Ferrari, D. W. Boukhvalov, M. I. Katsnelson, A. K. Geim, K. S. Novoselov, Science **323**, 610 (2009).

6. J. O. Sofo, A. S. Chaudhari, G. D. Barber, Phys. Rev. B **75**, 153401 (2007).

7. J. Zhou, M. Wu, X. Zhou, Q. Sun, Appl. Phys. Lett. **95**, 103108 (2009).

8. J. Zhou, Q. Wang, Q. Sun, X. S. Chen, Y. Kawazoe, P. Jena, Nano Lett. in press. http://pubs.acs.org/doi/pdfplus/10.1021/nl9020733.

9. J. P. Perdew, K. Burke, M. Ernzerhof, Phys. Rev. Lett. **77**, 3865 (1996).

10. G. Kresse, J. Furthmuller, Phys. Rev. B **54**, 11169 (1996).

11. G. Kresse, J. Joubert, Phys. Rev. B **59**, 1758 (1999).

12. H. J. Monkhorst, J. D. Pack, Phys. Rev. B **13**, 5188 (1976).

13. M. Topsakal, E. Akturk, S. Ciraci, Phys. Rev. B **79**, 115442 (2009).

14. S. H. Lai, K. L. Chang, H. C. Shih, K. P. Huang, P. Lin, Appl. Phys. Lett. **85**, 6248 (2004).

15. K. N. Kudin, H. F. Bettinger, G. E. Scuseria, Phys. Rev. B **63**, 045413 (2001).

16. K. H. An, J. G. Heo, K. G. Jeon, D. J. Bae, C. Jo,; C. W. Yang, C. Park, Y. H. Lee, Y. S. Lee, Y. S. Chung, Appl. Phys. Lett. **80**, 4235 (2002).

17. I. S. Neretin, K. A. Lyssenko, M. Y. Antipin, Y. L. Slovokhotov, O. V. Boltalina, P. A. Troshin, A. Y. Lukonin, L. N. Sidorov, R. Taylor, Angew. Chem. Int. Ed. **39**, 3273 (2000).

18. A. A. Popov, V. M. Senyavin, V. I. Keropanov, I. V. Goldt, A. M. Lebedev, V. G. Stankevich, K. A. Menshikov, N. Y. Svechnikov, O. V. Boltalina, I. E. Kareev, S.

Kimura, O. Sidorova, K. Kanno, I. Akimoto, Phys. Rev. B **79**, 045413 (2009).

19. Z. Slanina, F. Uhlik, Chem. Phys. Lett. **374**, 100 (2003).

20. A. A. Popov, V. M. Senyavin, O. V. Boltalina, K. Seppelt, J. Spandl, C. S. Feigerle, R. N. Compton, J. Phys. Chem. A **110**, 8645 (2006).

21. L. G. Bulusheva, A. V. Okotrub, O. V. Boltalina, J. Phys. Chem. A **103**, 9921 (1999).

22. S. I. Troyanov, P. A. Troshin, O. V. Boltalina, I. N. Ioffe, L. N. Sidonov,;E. Kemnitz, Angew. Chem. Int. Ed. **40**, 2285 (2001).

23. J. Jia, H. Wu, X. Xu, X. Zhang, H. Jiao, J. Am. Chem. Soc. **130**, 3985 (2008).

24. P. A. Troshin,; A. G. Avent,; A. D. Darwish,; N. Martsinovich,; A. K. Abdul-Sada,; J. M. Street,; R. Taylor, Science **309**, 278 (2005).

25. Z. Zhang, W. Guo, J. Am. Chem. Soc. **131**, 6874 (2009).

26. Q. Sun, Q. Wang, P. Jena, Y. Kawazoe, ACS Nano **2**, 341 (2007).

27. M. Wu, Q. Sun, Q. Wang, P. Jena, Y. Kawazoe, J. Chem. Phys. **130**, 184714 (2009).